\documentclass[useAMS,usenatbib]{mn2e}
\usepackage{amssymb,latexsym,amscd,amsmath,epsfig}

\def\be{\begin{equation}}
\def\ee{\end{equation}}
\def\ba{\begin{array}}
\def\ea{\end{array}}
\def\bea{\begin{eqnarray}}
\def\eea{\end{eqnarray}}
\def\bi{\begin{itemize}}
\def\ei{\end{itemize}}

\def\half{{\textstyle{1\over2}}}

\title{Upper limits on the observational effects of nuclear pasta in neutron stars}
\author[M. Gearheart et al]{M.~Gearheart, W.~G.~Newton, J.~Hooker and Bao-An~Li\\
Department of Physics and Astronomy, Texas A\&M University-Commerce, Commerce, Texas 75429-3011, USA}

\begin{document}

\date{\today}

\maketitle

\begin{abstract}

The effects of the existence of exotic nuclear shapes at the bottom of the neutron star inner crust - nuclear `pasta' - on observational phenomena are estimated by comparing the limiting cases that those phases have a vanishing shear modulus and that they have the shear modulus of a crystalline solid . We estimate the effect on torsional crustal vibrations and on the maximum quadrupole ellipticity sustainable by the crust. The crust composition and transition densities are calculated consistently with the global properties, using a liquid drop model with a bulk nuclear equation of state (EoS) which allows a systematic variation of the nuclear symmetry energy. The symmetry energy $J$ and its density dependence $L$ at  nuclear saturation density are the dominant nuclear inputs which determine the thickness of the crust, the range of densities at which pasta might appear, as well as global properties such as the radius and moment of inertia. We show the importance of calculating the global neutron star properties on the same footing as the crust EoS, and demonstrate that in the range of experimentally acceptable values of $L$, the pasta phase can alter the crust frequencies by up to a factor of three, exceeding the effects of superfluidity on the crust modes, and decrease the maximum quadrupole ellipticity sustainable by the crust by up to an order of magnitude. The signature of the pasta phases and the density dependence of the symmetry energy on the potential observables highlights the possibility of constraining the EoS of dense, neutron-rich matter and the properties of the pasta phases using astrophysical observations. 

\end{abstract}


\section{\label{sec1}Introduction}

The neutron star crust is expected to be for the most part an elastic solid, comprising a Coulomb lattice of nuclei, unbound electrons and, in the inner crust, fluid neutrons external to the nuclei \citep{BPS1971,BBP1971}. However, in the deep layers of the inner crust, near the crust-core boundary, the state of matter is more uncertain, and the possibility that the nuclei may undergo a series of shape transitions from spherical to cylindrical and slab, and then inverting to form the corresponding bubble configurations, has been widely studied \citep{Ravenhall1983,Hashimoto1984,Oyamatsu1984,Oyamatsu1993,Watanabe2001,Magierski2002,Gogelein2007}. More complex shapes can also exist \citep{Nakazato2009,Newton2009}. These phases are collectively termed `nuclear pasta'. Most studies have focussed on delineating the crustal region populated by the pasta phases. Important questions to address include: are there any practically observable neutron star phenomena that are sensitive to the properties of the pasta phase (assuming they are different to those of a Coulomb crystal)? In what aspects of neutron star modeling do we need to seriously pay attention to the properties of the pasta phases, and what modeling can we safely ignore the possibility of their existence?

The sensitivity of the pasta phases to neutrino interactions have been studied \citep{Horowitz2004,Horowitz2004_2,Gusakov2004}, and the possibility that the direct Urca cooling process is allowed in the bubble phases of the pasta introduces a potential observational probe should a young enough neutron star be found \citep{Gusakov2004}. The pasta phases are unlikely to be solid; they are strongly analogous to terrestrial complex fluids \citep{Pethick1998,Watanabe2005} which exhibit a wide range of responses to mechanical stimuli which are strongly dependent on the timescale over which the stimuli are applied. The modeling of mechanical properties such as the shear modulus is a very complex task, and requires simulating the structure and mechanical properties of matter over a wide range of length scales and timescales \citep{Pethick1998}. Usually, in modeling phenomena associated with elastic properties of the neutron star crust, the shear modulus is taken to be that of a Coulomb crystal over the whole crust.

In this paper we will estimate the maximal effect of the pasta phases on two elastic crustal processes that lead to potentially observable effects by comparing the case where the shear modulus in the pasta phases is taken to be that of a crystalline solid (which we refer to as `solid' pasta), with the case where the shear modulus is set to zero (which we we refer to as `liquid' pasta). The two neutron star properties we will examine are the maximal quadrupole deformation of the crust and the frequency of torsional oscillations of the crust. The former is important to determine the likelihood of observing gravitational wave emission mechanism from rotating, deformed neutron stars \citep{Abbott2010} and the latter is a proposed mechanism for generating quasi-periodic oscillations (QPOs) in the tails of light curves of giant flares from soft gamma-ray repeaters (SGRs) \citep{Israel2005, Watts2006, Strohmayer2005, Strohmayer2006}. 

An important ingredient in the crust and core models is the symmetry energy of nuclear matter as a function of density, $E_{\rm sym}(\rho)$, which can be defined as the energy difference per nucleon between pure neutron matter and symmetric nuclear matter (equal numbers of protons and neutrons) at a given density. Its density dependence $L$ at nuclear saturation density $n_0 \approx 0.16$ fm$^{-3}$, related to the pressure of pure neutron matter $p_0$ through the relation $L = p_0/ 3n_0$, is strongly correlated with the crust-core transition density and the thickness of the pasta layers \citep{Horowitz2001,Oyamatsu2007,Xu2008}, as well as global properties of the star such as the radius and moment of inertia \citep{Li2006,Worley2008}. We will use a simple crustal model which is self-consistent with the core equation of state, and vary the value of the slope of the nuclear symmetry energy at saturation density over its experimentally constrained range to take into account these correlations consistently. 

In section 2 we outline the approximate expressions we will use to estimate torsional frequencies and mountains, as well as give details about the EoS used and crustal model, including the shear modulus. In section 3 we give the results and in section 4 we discuss our conclusions.

\section{\label{sec2}Method}

The shear modulus of a Coulomb lattice of positively charged nuclei in a uniform negatively charged background in the neutron star crust at a baryon number density $n_{\rm b}$ was determined through Monte-Carlo simulation \citep{Ogata1990,Strohmayer1991,Chugunov2010} and can be written as

\be \label{eq:shear_mod}
	\mu = 0.1106 \left(\frac{4\pi}{3}\right)^{1/3} A^{-4/3} n_{\rm b}^{4/3} (1-X_{\rm n})^{4/3} (Ze)^2,
\ee

where the nuclei are characterized by the nucleon and proton number $A,Z$ and $X_{\rm n}$ is the fraction of neutrons not confined to the nuclei. This is the zero-temperature expression, a good approximation for neutron star temperatures below $\sim 10^8$K. We assume that the shear modulus is isotropic. From this one can calculate the speed of shear waves in the crust $v_{\rm s} = (\mu / \rho)^{\half}$, where $\rho$ is the mass density corresponding to the baryon number density $n_{\rm b}$.

We will use the expression to calculate the following quantities:

\begin{enumerate}
\item{The fundamental frequency and overtones of crustal torsional oscillations can be estimated as \citep{Samuelsson2007}

\be\label{freq}\begin{array}{l l}
	\displaystyle\omega^2_0 \approx \frac{e^{2\nu}v_s^2(l-1)(l+2)}{2RR_c},
\end{array}\ee

\be\label{over}
\omega^2_n \approx e^{\nu - \lambda} {n \pi v_s \over \Delta} \bigg[ 1 + e^{2\lambda} {(l-1)(l+2)\over 2 \pi^2} {\Delta^2 \over R R_c} {1 \over n^2} \bigg],
\ee

respectively, assuming an isotropic crust. $n$ is the number of radial nodes the mode has, and $l$ is the angular `quantum' number. In order to estimate the superfluid effects of the dripped neutrons, one introduces a multiplicative factor $\omega^2 \to \omega^2 \epsilon_{\star}$ \citep{Andersson2009}, where

\be \label{entrain}
\epsilon_{\star} = { (1 - X_{\rm n}) \over [1 - X_{\rm n} (m_{\rm n}^* / m_{\rm n})]}
\ee

encodes the effect of entrainment of the free neutrons by the lattice, $e^{2 \nu} = (1-{2GM}/{c^2R})$ and $R_{\rm c}$ is the radius of the crust-core boundary. This formula is derived from a plane wave analysis of the crustal shear perturbation equations \citep{Piro2005, Samuelsson2007}, and has, in a related form and without superfluid effects, been used to study the influence of the nuclear EoS on torsional mode frequencies \citep{SteinerWatts2009}. The mesoscopic effective mass $m_{\rm n}^*$ encodes the `drag' on the superfluid neutrons by the crystal lattice through which they flow; it has been calculated for a limited number of densities \citep{Chamel2005}. In this work we take the two extreme values $m_{\rm n}^*/m_{\rm n} = 15$ and $m_n^*/m_n = 1$, the latter corresponding to no entrainment effect. Note that the effective mass is the only quantity we do not calculate self-consistently with the crustal and core EoS used, something that should be addressed in future studies.}

\item{An estimate of the maximum quadrupole ellipticity sustainable by the crust of a neutron star of mass $M$ and radius $R$

\be \label{max_quad_def}
\epsilon \approx { \mu \over M/V_{\rm c} } \bigg( {GM \over R} \bigg)^{-1} \times \bar{\sigma}_{\rm max} \sim 10^{-7} \times \bigg( {\bar{\sigma}_{\rm max} \over 10^{-2} } \bigg),
\ee

where $V_{\rm c}$ is the crustal volume. This has the simple physical interpretation of the ratio the stress energy to the gravitational energy of the star multiplied by the breaking strain of the crust $\bar{\sigma}_{\rm max}$. This simple estimate has been confirmed by rigorous calculations \citep{Ushomirsky2000,Haskell2006}. The breaking strain has recently been estimated to be $\approx 0.1$ using molecular dynamics simulations \citep{Horowitz2009}.
Additionally, we calculate the equivalent gravitational wave strain amplitude from a rotating neutron star with a moment of inertia $I$, frequency $\nu$ and distance to Earth $d$ \citep{Abbott2007}:

\begin{align} \label{strain}
h_0 \approx  3.8 \times 10^{-25} & \bigg( {\epsilon \over 10^{-7} { \bar{\sigma}_{\rm max} \over 10^{-2}}  } \bigg) \bigg( {I \over 10^{45} {\rm g cm^2} } \bigg) \times \notag \\
                                                           & \bigg( {\nu \over 300 {\rm Hz}} \bigg)^2 \bigg( {d \over 0.1 {\rm kpc}} \bigg)^{-1}.
\end{align}}

\end{enumerate}

From our crustal model, we need the crustal composition and density at the crust-core and spherical-pasta transitions. We use the compressible liquid drop model (CLDM) developed in \citep{Newton2011}.  Like the liquid drop and droplet models of terrestrial nuclei, the CLDM focusses on the average, macroscopic properties of nuclear clusters in the crust such as their mass and size, whilst neglecting quantum shell effects. The interior nuclear density is a free variable (i.e. the nuclear clusters are compressible) rather than specified \emph{a priori} as the saturation density of symmetric nuclear matter. An external gas of dripped neutrons, expected to coexist with the nuclear clusters in the inner crust, is calculated using the same bulk Hamiltonian as the matter inside the clusters. The interfacial energy between the two bulk phases is obtained by specifying the magnitude and proton fraction dependence of the surface tension using the form of Lattimer \emph{et al} \citep{Ravenhall1983.2, Lattimer1985}. The neutron star inner crust is assumed to have a periodic structure microscopically, and the Wigner-Seitz (W-S) approximation is employed in which the unit cell, which generally will have a cubic or more complicated structure, can be replaced by a cell of the same volume and possessing the same geometry as the nuclear cluster. We consider spherical, cylindrical and planar geometries for both normal nuclear clusters and bubbles. The equilibrium cell size, nuclear size, proton fraction and dripped neutron density are calculated by minimizing the total energy of one cell at a given baryon number density. The energetically preferred phase is determined by calculating the equilibrium energy density for each of the six nuclear geometries and for uniform matter, and selecting the lowest. By repeating the calculation for the density range of the inner crust, we obtain the composition and the transition densities (the transition to pasta is taken to be the transition between spherical-cylindrical geometries). 

The bulk nuclear energy is given by a Skyrme-like interaction MSL \citep{MSL01}. The MSL interaction contains parameters directly related to the slope of the symmetry energy, allowing $L$ to be freely adjusted and giving us the ability to scan the appropriate parameter space of the nuclear interaction without changing the properties of the symmetric nuclear matter EoS. We take as a conservative range of $L$ from nuclear experiment $25 < L < 115 $ MeV \citep{Chen2005,Li2005,Shetty2007,Klimkiewicz2007, Danielewicz2007,Li2008,Centelles2009,Tsang2009,LieWenChen2010,ChangXu2010}, although it is worth noting that the most recent constraints \citep{Centelles2009,LieWenChen2010,ChangXu2010}, coupled with inference from theoretical calculations of pure neutron matter \citep{Hebeler2010,Gandolfi2011}, place the value of $L$ in the lower half of this range 25 - 75 MeV. We hold the absolute value of the nuclear symmetry energy at saturation density constant at $J \equiv E_{\rm sym}(n_0) = 32$ MeV throughout. 

We use the same uniform nuclear matter EoS as input into the Tolman-Oppenheimer-Volkoff equations to calculate the global neutron star properties needed (mass $M$, radius $R$, and crust thickness $\Delta = R - R_{\rm c}$, crust volume $V_{\rm c}$), and approximate the moment of inertia using the formula \citep{Lattimer2005}

\be \label{inertia}
I \approx 0.237 M R^2 \bigg[ 1 + 4.2 \bigg({M \over M_{\odot}}  { {\rm km} \over R }  \bigg) + 90 \bigg({M \over M_{\odot}}  { {\rm km} \over R }  \bigg)^4 \bigg]
\ee

\emph{As an upper estimate for the effect of nuclear pasta on the crustal shear phenomena, we will set the pasta shear modulus to zero}. We will thus be exploring two extreme cases distinguished by the density $n_{\rm t}$ at which we evaluate the shear modulus for use in Eqs. \ref{freq} and \ref{max_quad_def}:

\begin{enumerate}
\item{Pasta as an elastic solid; then the shear modulus is evaluated at the crust-core boundary $n_{\rm t} = n_{\rm crust \to core}$}
\item{Pasta as a liquid; then the shear modulus is evaluated at the boundary between the phase of spherical nuclei and the pasta phases $n_{\rm t} = n_{\rm sph \to pasta}$}
\end{enumerate}

Taking pasta phases to be liquid, we are effectively moving the boundary between the solid crust and the liquid core of the star down in density, so that the solid crust gets thinner. One would then expect the fundamental mode frequency to decrease, along with $\epsilon$. 

\section{\label{sec3}Results} 

Fig. 1 shows the density and pressure at the crust-core transition and the transition from spherical nuclei to the pasta phases. The difference between the two transitions indicate the range of density and pressure inhabited by the pasta phases, versus $L$. The well-known inverse correlation between the crust-core transition density and $L$ is manifest; the corresponding inverse correlation between the crust-core transition pressure and $L$ is also shown, although one should note that this relation is dependent on the nuclear model used \citep{Ducoin2011} and is sensitive to other parameters of the nuclear EoS. The density range of pasta is largest at the smallest values of $L$, and disappears at the high end of the experimentally constrained range of $L$.

\begin{figure}\label{fig:1}
\begin{center}
\includegraphics[width=55mm,height=80mm]{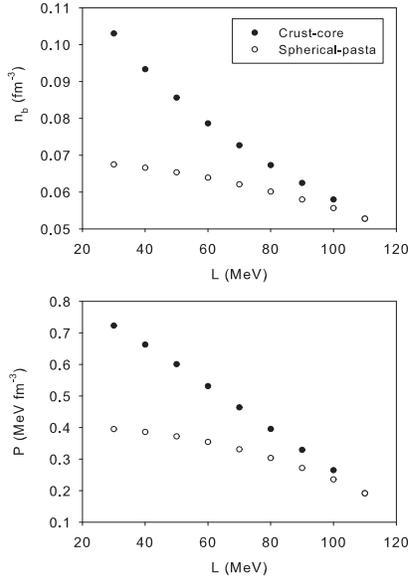}
\caption{The transition density (top) and pressure (bottom) as a function of the slope of the nuclear symmetry energy $L$. Filled circles refer to the crust-core transition and empty circles show the transition between spherical nuclei and the pasta phases in the inner crust.}
\end{center}
\end{figure}

Fig. 2 shows the crust composition parameters necessary for the determination of the shear modulus in a Coulomb crystal: the total nucleon number of the nuclear cluster $A$, the corresponding proton number $Z$ and the fraction of dripped neutrons in the unit cell $X_{\rm n}$. $A$ and $Z$ at the crust-core transition generally decrease as $L$ increases, a trend that is largely due to the fact that the crust-core transition density also decreases with $L$ (and $A$ and $Z$ rise with density at the highest densities in the inner crust). At the spherical nuclei - pasta transition, $A$ is reasonably uniform, while $Z$ still decreases from 30 to about 10. It is important to note that in the semi-classical liquid drop model, shell effects are not included, which could significantly affect the trend. The dripped neutron fraction $X_{\rm n}$ is larger at the spherical nuclei - pasta transition (as is the dripped neutron mesoscopic effective mass \citep{Carter2005}); $X_{\rm n}$ increases with $L$ at both transition regions.

\begin{figure}\label{fig:2}
\begin{center}
\includegraphics[width=55mm,height=100mm]{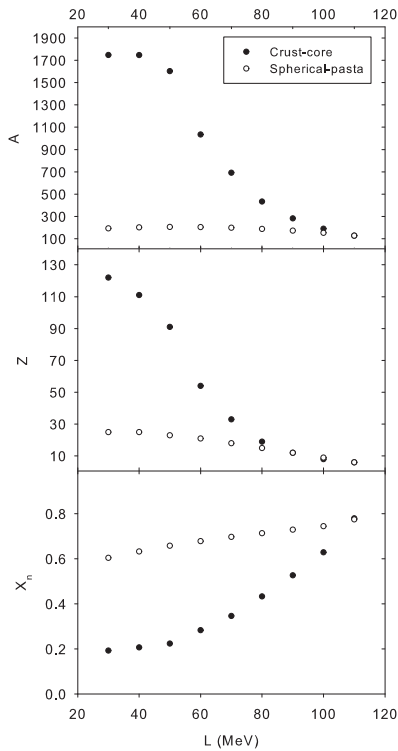}
\caption{Composition of the bottom of the inner crust: the total nucleon number (top), proton number (middle) and global dripped neutron fraction (bottom) at the two different transition densities referred to in Fig. 1, as a function of the slope of the nuclear symmetry energy $L$.}
\end{center}
\end{figure}

Fig. 3 shows the shear modulus relative to the pressure, and the shear speed, at the two transition densities as a function of $L$, using the compositional and transition parameters from figs. 1 and 2. It is immediately obvious that the influence of the pasta phases disappears at $L \gtrsim 80$ MeV. The shear modulus and speed decrease as $L$ increases; thus shear phenomena in the crust will be strongest for $L$ at the lower end of the experimentally constrained range.

\begin{figure}\label{fig:3}
\begin{center}
\includegraphics[width=55mm,height=80mm]{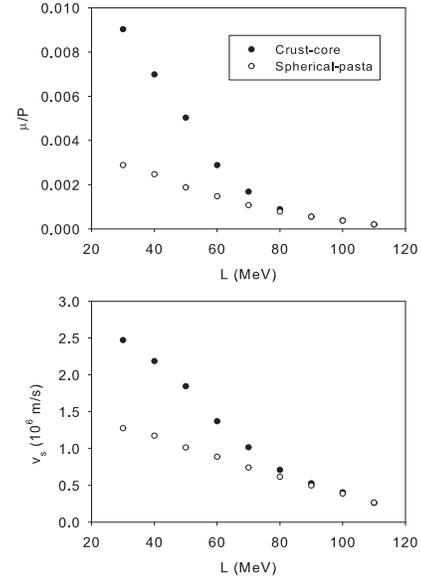}
\caption{The shear modulus compared with the pressure (top) and the shear speed (bottom) at the two different transition densities referred to in Fig. 1, as a function of the slope of the nuclear symmetry energy $L$.}
\end{center}
\end{figure}

Figs 4 and 5 show the relevant global neutron star and neutron star crust properties for a $1.4 M_{\odot}$ star. The radius $R$ increases as $L$ increases, as one would expect: since $L$ at nuclear saturation density is a good fiducial indicator of the internal neutron star pressure, one would expect that the higher the pressure, the larger the radius. This is a relation that is well known \citep{Lattimer2001,Li2006}. The thickness and volume of the crust that contains spherical nuclei and the total crust with the pasta phases included both increase with $L$ at $L \lesssim 80$ MeV and then decrease at higher $L$. This trend is the convolution of the decrease of the transition densities with $L$ and the increase of the stellar radius with $L$.  The moment of inertia follows a similar trend. The thickness of the pasta layer is greatest at intermediate values of $L \approx 70$ MeV, rising to about 10\% of the crustal thickness and volume. It is important to note that, although the pasta layers occupy a small layer of the crust in terms of radius of volume, they occupy the highest density region of the crust, so the fraction of the mass of the crust occupied by the pasta layers will be much higher, approaching 50\% \citep{Lorenz1993}.

\begin{figure}\label{fig:4}
\begin{center}
\includegraphics[width=55mm,height=75mm]{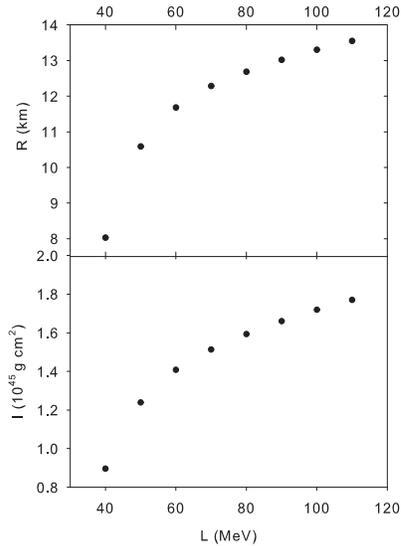}
\caption{Global neutron star properties as a function of the slope of the nuclear symmetry energy $L$: for a 1.4$M_{\odot}$ neutron star, the radius (top), and moment of inertia (bottom) are plotted.}
\end{center}
\end{figure}

\begin{figure}\label{fig:5}
\begin{center}
\includegraphics[width=55mm,height=75mm]{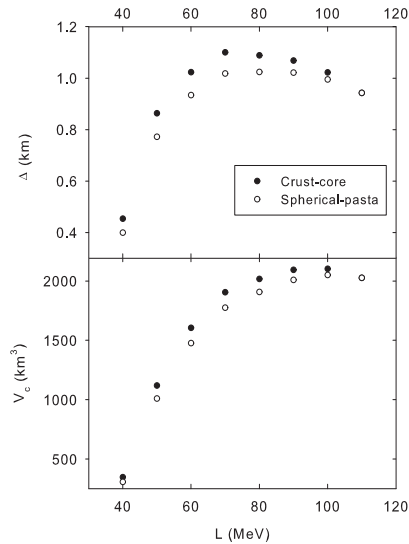}
\caption{Global crust properties as a function of the slope of the nuclear symmetry energy $L$: for a 1.4$M_{\odot}$ neutron star, the crust thickness (top) and crust volume (bottom) are plotted. The latter two quantities are shown taking the bottom of the crust to be the true crust-core boundary (filled circles) and the spherical nuclei-pasta boundary (empty circles).}
\end{center}
\end{figure}

Given the microscopic and global neutron star properties, both consistently calculated with the same underlying neutron star EoS, we can now estimate the maximum quadrupole ellipticity sustainable by the crust and the fundamental mode frequency of torsional crust oscillations. Fig. 6 shows the maximum quadrupole ellipticity of the crust of a $1.4 M_{\odot}$ star, normalized to the canonical value of $10^{-7} (\bar{\sigma}_{\rm break}/10^{-2})$, as a function of $L$ with the pasta phases taken to be a Coulomb solid and a liquid. The maximum quadrupole ellipticity peaks at $L \approx 50$ MeV; the peak occurs because of the competing trends of increasing stellar radius and decreasing crust-core transition density as $L$ increases, a feature which only emerges when the crust and core are treated consistently. For $L \lesssim 70$ MeV, and taking the pasta phases to be solid, the maximum ellipticity is close to the canonical value. However, if the pasta phases are assumed to be liquid, the maximum ellipticity is an order of magnitude lower $\sim 10^{-8} (\bar{\sigma}_{\rm break}/10^{-2})$. At higher values of $L$, the pasta phases become insignificant, but the stiffening of the EoS and reduction of the crust-core transition density means that the crust thickness relative to the size of the star overall becomes smaller, and the maximum ellipticity sustainable falls by almost two orders of magnitude compared to the canonical value. The equivalent gravitational wave strain is plotted in the bottom half of Fig. 6 for a reference neutron star frequency of 300Hz and distance of 0.1kpc. The sensitivity of the most recent LIGO science run to such gravitational waves is indicated by the dashed line; this shows that, despite the reduced strain caused by a stiff EoS or the existence of liquid pasta phases, such GWs are still potentially observable with current or future detectors, offering the chance to distinguish between the various possibilities of crust properties. 

\begin{figure}\label{fig:6}
\begin{center}
\includegraphics[width=60mm,height=80mm]{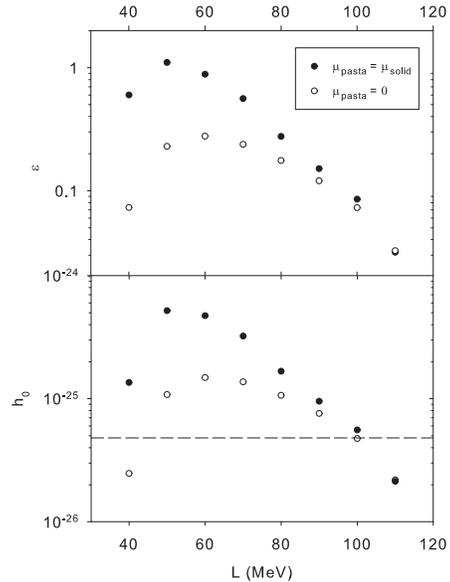}
\caption{Estimate of the maximum quadrupole ellipticity of a 1.4$M_{\odot}$ neutron star (top) and the corresponding gravitational wave strain (bottom) as a function of the slope of the nuclear symmetry energy $L$. The strain is calculated assuming a neutron star frequency of 300Hz and a distance of 0.1kpc from Earth. The filled circles indicate the value taking the shear modulus in the pasta phases to be that of a solid Coulomb lattice; the empty circles indicate the value taking the shear modulus in the pasta phase to be zero (i.e. liquid pasta). The dashed line in the bottom plot indicates the sensitivity of the most recent LIGO science run \citep{Abbott2010}.}
\end{center}
\end{figure}

Fig. 7 shows the fundamental frequency $l=2, n=0$ and first overtone $l=2, n=1$ of the crustal torsional oscillation spectrum for a $1.4 M_{\odot}$ star, where $n$ is the number of radial nodes and $l$ the angular constant associated with the spherical harmonics $Y_l^m$. Observed values of the frequency of QPOs at 16, 26, 28 and 30 Hz \citep{Israel2005, Watts2006, Strohmayer2005, Strohmayer2006}, within the range of the calculated fundamental frequencies, are indicated by the dashed lines, as well as the observed 84, 92, 150, 155 and 625 Hz frequencies within the range of the first overtone. The frequency generically decreases with increasing $L$ as the crust becomes thinner relative to the radius of the star. Ignoring superfluid effects and the effects of the pasta phases, the frequency matches observed QPOs from SGRs only at the lowest values of $L$. If the pasta phases are liquid, the frequency falls by a factor of 3, making it difficult to match the 28Hz frequency observed, and being consistent with the 18Hz observed frequency only at the lowest value of $L$. Compare to the effects of superfluidity, the effects of \emph{liquid} pasta are about a factor of 4 larger. It should be noted that the observed frequencies could also be explained by core Alfv\'en waves \citep{Sotani2008}. $n=0$ modes scale with $[(l+2)(l-1)]^{1/2}$, so higher-$l$ modes will show a similar effect of the pasta phases. The same trend is present for the first overtone. It is interesting to note that the observed 625Hz mode, often cited as the main candidate for the first overtone, is consistent only with low $L < 50$ MeV and solid pasta, whereas the lower observed frequencies 84-155Hz, which are often matched with $n=0$, $l>2$ modes, are consistent with a wider range of $L$, 60-110 MeV, and pasta properties.

\begin{figure}\label{fig:7}
\begin{center}
\includegraphics[width=60mm,height=80mm]{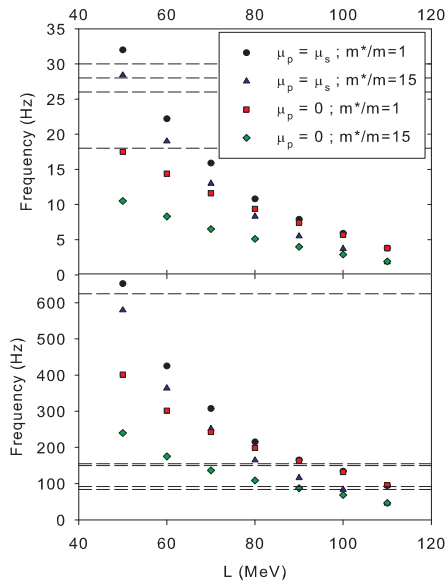}
\caption{(Color online) The frequency of the fundamental torsional oscillation mode (top) and the first overtone (bottom) in the the crust for a 1.4$M_{\odot}$ as a function of the slope of the nuclear symmetry energy $L$. The circles and triangles show the frequency assuming the shear modulus in the pasta phases to be that of a Coulomb lattice; the triangles take into account the entrainment of the superfluid neutrons by the nuclear clusters. The squares and diamonds show the frequency assuming the shear modulus in the pasta phase to be zero (i.e. liquid pasta); the diamonds take into account the entrainment of the superfluid neutrons by the nuclear clusters. The dashed lines show possible candidate frequencies for the fundamental modes; 18, 26, 28, 30Hz in the fundamental frequency range and 84, 92, 150, 155Hz in the range of the first overtone.}
\end{center}
\end{figure}

\section{\label{sec4}Conclusions}

We have demonstrated that the pasta phases can have a very significant effect on observable neutron star phenomena, reducing the frequencies of crustal torsional modes by up to a factor of 3 and the maximum quadrupole ellipticity sustainable by the crust by an order of magnitude. The effect of the pasta phases is comparable with other transport properties such as entrainment of superfluid neutrons. In addition, we have demonstrated that a consistent treatment of the crust composition and core equation of state is required in modeling these phenomena, and that when one uses such a consistent treatment the possibility of constraining the nuclear symmetry energy at densities around saturation using astrophysical observations emerges. The models of torsional modes and mountain building in neutron star crusts have many other uncertainties. Having a diverse range of neutron star phenomena amenable to independent observation allows one to check the consistency of our models; studying the signature of nuclear matter properties in such phenomena opens up another set of constraints from nuclear experiment. 

Clearly, more work needs to be done investigating the likely structure of the pasta phases over a variety of length scales and estimating their actual shear modulus. This is likely to depend on the particular shapes present. One also cannot ignore temperature effects, especially for the long timescale process of accreting material onto the crust \citep{Chugunov2010}. The pasta phases are likely to be disordered on length scales of $\approx 100$ cell lengths due to thermal fluctuations \citep{Watanabe2000}; however, we cannot rule out some ordering mechanism involving, for example, the magnetic field. One might envisage the cylinders or slabs aligning along a particular direction, giving the phases liquid crystal-like properties \citep{Pethick1998}. The approximation of pasta being a liquid may be quite good for accretion processes which build mountains, as it is known that many complex fluids exhibit continuous flow in response to stress applied slowly over long timescales. A more microscopic treatment of crustal matter, including shell effects of nuclei and dripped neutrons, is also necessary for a more realistic determination of the transition densities and superfluid effects. We hope that demonstrating the feasibility of setting limits on those properties observationally provides additional motivation for these further studies.

While finalizing this manuscript, we learned of a similar study which focusses on the torsional crust oscillations \citep{Sotani2011}. We briefly comment on how our study compares with the results obtained there. The main difference is that, whereas we use approximate methods for the description of the torsional modes and a consistent treatment of crust and core EoS, \citet{Sotani2011} uses a more sophisticated calculation of the crustal frequencies, but a specific crust and core EoS which are not consistent with each other. The core EoS used in \citet{Sotani2011} is relatively stiff, giving neutron star radii consistent in our model with $L > 100$ MeV \citep{Sotani2008}, putting it outside of current best experimental and observational estimates \citep{Steiner2010}. The crustal composition model is that of Douchin and Haensel (DH) \citep{Douchin2001}, using the SLy4 EoS with $L = 46$ MeV. The transition density to the pasta phases is taken as a free parameter rather than emerging consistently from the crustal model (indeed, the DH EoS predicts no pasta phases); the lower value taken, $n_{\rm t} \approx 0.007$ fm$^{-3}$, is well below that which is realistically expected from liquid drop and more microscopic models. The range we find is 0.05 - 0.07 fm$^{-3} \equiv 7 \times 10^{13} - 10^{14}$ g cm$^{-3}$. The differences in our methods makes a direct comparison of our results difficult. For example, \citet{Sotani2011} obtains a fundamental mode frequency of $\approx 22$ Hz for a $1.4 M_{\odot}$ neutron star when the shear modulus is taken to be that of a Coulomb crystal up to the crust-core boundary, compared with $\approx 34$ Hz in our model, taken at $L = 46$ MeV; however, the stiffness of the crust in the Sotani model would lead to a thinner crust and a decrease in the frequency compared with our results. It would be useful to make a comparison of the methods used using consistent nuclear physics. As a final note, we have estimated the comparative effect of the pasta phases and the effects of superfluid neutrons, finding them to be comparable, as suggested in \citet{Sotani2011}.

\section*{ACKNOWLEDGEMENTS}
This work is supported in part by the National Aeronautics and Space Administration under grant NNX11AC41G issued through the Science Mission Directorate and the National Science Foundation under grants PHY-0757839 and PHY-1068022 and the Texas Coordinating Board of Higher Education under grant No. 003565-0004-2007.

\end{document}